\documentclass[a4paper,twocolumn,superscriptaddress,showpacs]{revtex4}
\usepackage{dcolumn}
\usepackage{amsmath}
\usepackage{graphicx}
\usepackage{latexsym}  
\usepackage{amsfonts}  
\usepackage{amssymb}  
\usepackage{color}
\DeclareGraphicsExtensions{.eps,.pdf,.png,.gif,.jpg}  
  
\newcommand{\be}{\begin{equation}}
\newcommand{\beq}{\begin{equation}}
\newcommand{\ee}{\end{equation}}
\newcommand{\bea}{\begin{eqnarray}}
\newcommand{\eea}{\end{eqnarray}}
\newcommand{\g}{\gamma}
\renewcommand{\vr} {{\bf r}}
\newcommand{\vg} {{\bf g}}
\newcommand{\vH} {{\bf H}}
\newcommand{\vA} {{\bf A}}
\newcommand{\vSi} {{\bf \Sigma}}
\newcommand{\nn} {\nonumber}

\begin{document}

\title{Time-dependent bond-current functional theory for lattice 
Hamiltonians: fundamental 
theorem and application to electron transport} 
 
\author{S. Kurth}  
\affiliation{Nano-Bio Spectroscopy Group, 
Dpto. de F\'{i}sica de Materiales, 
Universidad del Pa\'{i}s Vasco UPV/EHU, Centro F\'{i}sica de Materiales 
CSIC-UPV/EHU, Av. Tolosa 72, E-20018 San Sebasti\'{a}n, Spain} 
\affiliation{IKERBASQUE, Basque Foundation for Science, E-48011 Bilbao, Spain}
\affiliation{European Theoretical Spectroscopy Facility (ETSF)}
  
\author{G. Stefanucci}
\affiliation{Dipartimento di Fisica, Universit\`{a} di Roma Tor Vergata,
Via della Ricerca Scientifica 1, 00133 Rome, Italy}
\affiliation{European Theoretical Spectroscopy Facility (ETSF)}
  
\date{\today}  

\begin{abstract}
The cornerstone of time-dependent (TD) density functional 
theory (DFT), the Runge-Gross theorem, proves a 
one-to-one correspondence between TD potentials and TD densities of 
{\em continuum}  Hamiltonians. In all practical implementations, however, the 
basis set is {\em discrete} and  the system is effectively described by a 
lattice Hamiltonian. 
We point out the difficulties of 
generalizing the Runge-Groos proof to the discrete case and thereby 
endorse the recently proposed 
TD bond-current functional theory (BCFT) as a viable alternative.
TDBCFT is based on a one-to-one correspondence 
between TD Peierl's phases and TD bond-currents of lattice systems.    
We apply the TDBCFT formalism to electronic transport through 
a simple interacting device 
weakly coupled 
to two biased non-interacting leads. 
We employ Kohn-Sham Peierl’s phases which are discontinuous functions of
the density, a crucial property to describe Coulomb blockade. 
As shown by explicit time propagations, the discontinuity 
may prevent the biased system from ever reaching a steady state.
\end{abstract}
  
\pacs{31.15.ee, 73.23.Hk, 05.60.Gg}
  
\maketitle  

\section{Introduction}

The central idea of time-dependent (TD) density functional theory (DFT) is to map 
a time-dependent and interacting many-particle system onto a time-dependent 
system of non-interacting particles moving in an effective Kohn-Sham 
(KS) potential, $v_{\rm KS}(\vr,t)$, chosen such that the TD densities of the 
interacting and non-interacting systems are equal. From the knowledge 
of $v_{\rm KS}(\vr,t)$ it is then possible to compute the TD density $n(\vr,t)$, and hence 
the TD longitudinal current, in a one-particle manner.
With this results as a starting point, an exact TDDFT formulation of quantum 
transport which can cope with both transient and steady-state regimes
 has been proposed \cite{StefanucciAlmbladh:04,KurthStefanucciAlmbladhRubioGross:05}.

The theoretical foundation for the aforementioned mapping is laid down by the celebrated Runge-Gross theorem 
\cite{RungeGross:84} whose essential statement is that the time evolution 
of two systems evolving from the same initial state $|\Psi_0 \rangle$ under 
the influence of two different TD potentials $v(\vr,t)$ and 
$v'(\vr,t)$, which are analytic in time and differ by more than a purely 
time-dependent, position-independent function $C(t)$, leads to different 
TD densities $n(\vr,t)$ and $n'(\vr,t)$. This theorem was 
developed further by van Leeuwen \cite{Leeuwen:99} who showed that 
for a many-body system with a given particle-particle interaction 
$w(|\vr-\vr'|)$  and moving in 
some TD potential $v(\vr,t)$, 
the TD density can be reproduced in another 
many-body system with a {\em different} interaction $w'(|\vr-\vr'|)$ 
and moving in a TD potential $v'(\vr,t)$ provided that the initial 
states $| \Psi(0) \rangle$ and $| \Psi'(0) \rangle$ of the two systems yield 
the same density and longitudinal current. Moreover, for a given initial state the potential 
$v'(\vr,t)$ is unique up to a purely time-dependent function. 
Recently, Ruggenthaler and van Leeuwen 
extended the validity of these results by relaxing the condition of 
Taylor-expandability in time \cite{rvl.2010}. 

The van Leeuwen theorem reduces to the Runge-Gross theorem for $w=w'$ 
and $| \Psi(0) \rangle=| \Psi'(0) \rangle$, and establishes the 
existence of a non-interacting KS system  for $w'=0$ and $| \Psi'(0) 
\rangle=|\Phi_{\rm KS}\rangle$ 
a Slater determinant. In this latter case the 
potential $v'(\vr,t)$ reproducing a given density becomes the Kohn-Sham (KS)  
potential $v_{\rm KS}(\vr,t)$ of TDDFT.  In general, the KS potential is 
uniquely determined by the time-dependent density $n(\vr,t)$, and the initial 
states $|\Psi(0)\rangle$ and $|\Phi_{\rm KS}\rangle$. 
If we further assume that the initial states $|\Psi(0)\rangle$ and 
$|\Phi_{\rm KS}\rangle$ are ground states, then via the Hohenberg-Kohn 
\cite{HohenbergKohn:64} and Kohn-Sham \cite{KohnSham:65} theorems of static 
density functional theory both initial states are uniquely determined by the 
initial density $n(\vr,0)$, and the KS potential $v_{\rm KS}(\vr,t)$ becomes a unique 
functional of the density alone. The TD density can then 
be calculated from the TD KS equations \cite{RungeGross:84} 
\be
i \frac{\partial}{\partial t} \varphi_j(\vr,t) = \left( - \frac{\nabla^2}{2} 
+ v_{\rm KS}[n](\vr,t) \right) \varphi_j(\vr,t), 
\ee 
and 
\be
n(\vr,t) = \sum_j^{\rm occ} | \varphi_j(\vr,t) |^2,
\label{tddens}
\ee
where the sum in Eq.~(\ref{tddens}) is over the occupied orbitals in the KS 
Slater determinant. 

Generally, in practical implementations of the continuum TD KS equations there 
are three independent sources of errors: (i) the use of only a finite 
number of basis functions, i.e., a {\em de facto} description of the problem 
by a discrete or lattice Hamiltonian; (ii) the approximation 
employed for the exchange-correlation (XC) potential and (iii) inherent 
numerical inaccuracies. Even if we could use the exact KS potential 
(designed for continuum systems) and run the program on a machine with infinite 
accuracy, when using a finite basis set the densities of the 
KS system and of the original interacting system will, in general, differ. 
In fact, it would be 
useful to have a discrete version of TDDFT for ruling out 
the errors due to the incompleteness of the basis set. An even 
stronger motivation 
for constructing a discrete TDDFT is to have an 
alternative tool for the study of model systems like, e.g., the Anderson 
model, the Hubbard model, etc. Recent works have shown that not all densities 
can be represented by lattice Hamiltonians \cite{b.2008}. The breakdown of 
this $v$-representability is due to the fact that the rate of density change 
which can be supported on a given lattice is limited by the inverse of the 
hopping matrix elements \cite{LiUllrich:08,Verdozzi:08}.
In the next section we point out an even more severe problem related to the 
difficulties of generalizing the Runge-Gross proof to discrete or lattice 
Hamiltonians and re-understand why the continuum formulation works from 
the ``discrete perspective''. To overcome 
these difficulties a TD bond-current 
functional theory (BCFT) has 
recently been proposed \cite{StefanucciPerfettoCini:10}; here the 
underlying  
idea is to use the so called {\em Peierls phases} (discrete version 
of the vector potential)  instead of the scalar potential as the basic KS field. We  will
present the TDBCFT and demonstrate the analogous of the van 
Leeuwen theorem in discrete systems. In Section \ref{modhamsec} we 
take advantage of the 
TDBCFT formulation to construct a TD framework for quantum 
transport in model Hamiltonians. As an application in Section 
\ref{CBsec} we will revisit the 
Coulomb Blockade (CB) phenomenon in the light of a recent finding 
on the relevance of  the derivative discontinuity of static 
DFT functionals \cite{PerdewParrLevyBalduz:82} in the CB regime 
\cite{KurthStefanucciKhosraviVerdozziGross:10}.   
Conclusions and future perspectives are drawn in Section 
\ref{concsec}.

\section{Time-dependent density functional theory on a lattice}

\subsection{Problems in generalizing Runge-Gross proof}

We write the Hamiltonian of interacting electrons in some 
orthonormal basis $\varphi_m(\vr)$ of orbitals localized  
at site $m$ by expanding the field operators as $\hat{\psi}_{\sigma}(\vr) = \sum_m 
\hat{c}_{m \sigma}\varphi_m(\vr)$, where the creation (annihilation) 
operators $\hat{c}_{m \sigma}^{\dagger}$ ($\hat{c}_{m \sigma}$) satisfy the 
anticommutation relations 
\be
\left\{ \hat{c}_{m \sigma} , \hat{c}_{n \sigma'}^{\dagger} \right\} = 
\delta_{nm} \delta_{\sigma \sigma'} \; .
\ee
For electrons exposed to a TD on-site potential $v_m(t)$, the 
Hamiltonian then reads
\be
\hat{H}(t) = \hat{K}_0 + \hat{V}(t) + \hat{H}_{\rm int},
\ee
where the non-interacting part consists of a static part
\be
\hat{K}_0 = \sum_{m,n} \sum_{\sigma} T_{mn} \hat{c}_{m \sigma}^{\dagger} 
\hat{c}_{n \sigma} 
\label{niham_1}
\ee
and a time-dependent external potential
\be
\hat{V}(t) = \sum_m \sum_{\sigma} v_m(t) \hat{c}_{m \sigma}^{\dagger} 
\hat{c}_{m \sigma} \; .
\label{niham_2}
\ee
where we assume that $v_m(t)=0$ for all times $t<0$. 
$\hat{H}_{\rm int}$ denotes the operator 
describing the electron-electron interaction. Note that $v_{m}(t)$ is coupled 
to the density operator 
$\hat{n}_{m}=\sum_{\sigma}\hat{c}_{n \sigma}^{\dagger} \hat{c}_{n \sigma}$ 
at site $m$.  

In order to have a theoretical foundation for TDDFT on a lattice, one first 
has to prove the existence of a unique mapping between the 
expectation value $n_m(t)$ of the density operator at site $m$
and the on-site potential $v_m(t)$. The obvious idea 
is to adapt the original Runge-Gross proof \cite{RungeGross:84} for continuum 
Hamiltonians to the case of lattice Hamiltonians. However, this strategy runs 
into problems as we will now demonstrate. 

\begin{figure}[t]
\includegraphics[scale=.3,angle=-90,keepaspectratio=true]{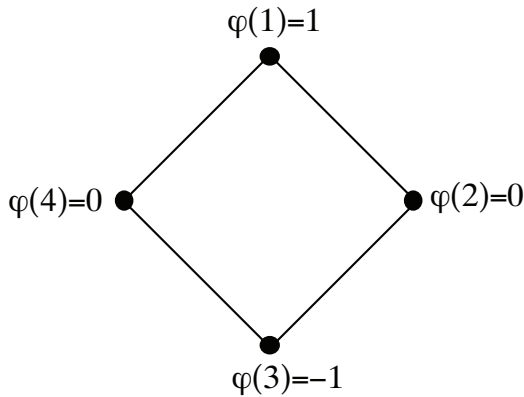}
\caption{The 4-site ring described in the main text 
with a single-particle eigenstate of energy zero.  Any external potential $v_{m}(t)$ with 
$v_{1}(t)=v_{3}(t)=0$ leaves the density unchanged. }
\label{counterexample}
\end{figure}

Let $| \Psi(t) \rangle$ be the many-body state that solves  
the TD Schr\"odinger equation 
\be
 i \frac{d}{d t} | \Psi(t) \rangle = \hat{H}(t) | \Psi(t) \rangle 
\ee
with the initial condition $| \Psi(t=0) \rangle = | \Psi_0 \rangle$. 
Then the expectation value $O(t) = \langle \hat{O}(t) \rangle = 
\langle \Psi(t)| \hat{O}(t) |\Psi(t) \rangle$ of 
any quantum mechanical operator $\hat{O}(t)$ satisfies the equation of motion 
\be
\frac{\rm d}{{\rm d} t} O(t) = \left\langle \frac{\partial \hat{O}(t)}
{\partial t} \right\rangle + i \left\langle [ \hat{H}(t),\hat{O}(t) ] 
\right\rangle \; .
\label{heseq}
\ee
For a straightforward generalization of the Runge-Gross proof to lattice 
systems one needs to show that two wavefunctions $|\Psi(t)\rangle$ and 
$|\Psi'(t)\rangle$ which evolve under the influence of two different 
potentials $v_m(t) \neq v'_m(t)$ from the {\em same} initial state 
$|\Psi(0)\rangle = |\Psi'(0)\rangle = |\Psi_0\rangle$ always lead to 
different TD densities $n_m(t)\neq n'_m(t)$. Let us, for simplicity, 
consider the simplest case of 
two time-independent potentials $v_{m}$ and $v'_{m}$ that differ by 
more than an additive constant. To show that they generate two 
different TD densities we calculate the succesive time-derivatives in 
$t=0$ of $n_m(t)$ and $n'_m(t)$ until the order (if any) at which they differ.
The zero-th and first derivatives in $t=0$ are clearly the same. From 
Eq. (\ref{heseq}) the second derivative of $n_m(t)$ and $n'_m(t)$ in 
$t=0$ reads
\be
\ddot{n}_{m}(0)=\sum_{k}(T_{mk}\rho_{km}+\rho_{mk}T_{km})(v_{k}-v_{m})+\Delta_{m},
\ee
\be
\ddot{n}'_{m}(0)=\sum_{k}(T_{mk}\rho_{km}+\rho_{mk}T_{km})(v'_{k}-v'_{m})+\Delta_{m},
\ee
with 
$\rho_{mk}=\sum_{\sigma}\langle\Psi_{0}|\hat{c}^{\dagger}_{k\sigma}\hat{c}_{m\sigma}|\Psi_{0}\rangle$
the one-particle density matrix at $t=0$ and
\bea
\Delta_{m} &=& - \Big\langle \Psi_0 \Big| \, \left[ \hat{K}_0,
[\hat{K}_0,\hat{n}_m ]\right]\Big|\Psi_0\Big\rangle \nn \\
&&- \Big\langle\Psi_{0}\Big|\left[\hat{H}_{\rm int},
[\hat{H}_{\rm int},\hat{n}_{m}]\right]\Big|\Psi_{0}\Big\rangle \;.
\eea
Introducing the short-hand notation 
$K_{mk}=T_{mk}\rho_{km}+\rho_{mk}T_{km}$ for the kinetic energy 
density of the bond $k-m$, the difference $\delta \ddot{n}=\ddot{n}(0)-\ddot{n}'(0)$ is 
 simply given by the formula below
\be
\delta \ddot{n}_{m}=\sum_{k}K_{mk}(\delta 
v_{k}-\delta v_{m}),
\label{2ndderdiff}
\ee
where $\delta v=v-v'$. 
Like in the original Runge-Gross proof we now proceed by {\em reductio ad 
absurdum}. Assume that $\delta \ddot{n}$ vanishes for a non-vanishing 
(and non-constant) $\delta v$. Then, multiplying both sides of 
Eq. (\ref{2ndderdiff}) by $\delta v_{m}$ and summing over $m$ we find
\be
0=\frac{1}{2}\sum_{km}K_{mk}(\delta 
v_{k}-\delta v_{m})^{2},
\label{noabseq}
\ee
where we made use of the symmetry $K_{mk}=K_{km}$.
Due to the fact that the $K_{mk}$ do not have a definite sign Eq. 
(\ref{noabseq}) is not an {\em absurdum}. In fact, in a lattice 
theory we could have two different potentials that yield the same 
second derivative of the density in $t=0$. 
We should then proceed further and try to prove that the third 
derivatives are different. One immediately realizes, however, that 
this way does not lead anywhere. There exist several counterexamples 
to the existence of a one-to-one correspondence between densities and 
potentials in a lattice 
Hamiltonian.  Consider, for example, a one-dimensional ring with 
four sites, zero on-site energies $T_{mm}=0$ 
and nearest-neighbor hopping 
$T_{12}=T_{23}+T_{34}=T_{41}=T$, see Fig. \ref{counterexample}. 
The single-particle state with 
amplitudes $\varphi(1)=1$, $\varphi(3)=-1$ and $\varphi(2)=\varphi(4)=0$ is an 
eigenstate with energy zero. Any external potential $v_{m}(t)$ with 
$v_{1}(t)=v_{3}(t)=0$ leaves the density unchanged.

In the next Section we show how to overcome these difficulties by 
considering the Peierls phases instead of the 
on-site potentials as the basic KS fields. Before, however, we find 
particularly instructive to re-understand why the continuum 
formulation works starting from the ``discrete'' perspective.
To limit the complications we will focus on one-dimensional systems.
Let $m=0,\pm 1,\pm 2,\ldots$ be the label of a grid point $x_m = m \Delta_x$ 
of a one-dimensional grid with lattice spacing $\Delta_{x}$.
Choosing the $T_{mn}$ such that it gives the three-point 
discretization of the (continuum) kinetic energy
\be
T_{mn}=-\frac{1}{2 \Delta_{x}^{2}}
\left(\delta_{m,n+1}-2\delta_{m,n}+\delta_{m,n-1}\right),
\label{tmnlap}
\ee
it is easy to show that the non-interacting Hamiltonian 
$\hat{H}_0(t) = \hat{K}_0 + \hat{V}(t)$ (see Eqs.~(\ref{niham_1}) and 
(\ref{niham_2})) approaches the continuum Hamiltonian
\be
\lim_{\Delta_{x}\to 0}\hat{H}_{0}(t)=\int dx \;
\hat{\psi}^{\dag}(x)\left[-\frac{\nabla^{2}}{2}+v(x,t)\right]\hat{\psi}(x) ,
\ee
where we defined the field operators
\be
\hat{\psi}(x_m)=
\lim_{\Delta_{x}\rightarrow 0}\frac{\hat{c}_{m}}{\sqrt{\Delta_{x}}}\;,
\ee
and the scalar potential $v(x_m,t)= v_{m}(t)$. 
Inserting the $T_{mn}$ of Eq. (\ref{tmnlap}) into Eq. 
(\ref{noabseq}) and taking the limit $\Delta_{x}\to 0$ we find
\bea
0&=&\lim_{\Delta_{x}\to 0}\Delta_{x}\sum_{k}{\rm Re}\left[\frac{\rho_{kk+1}}{\Delta_{x}}\right]\left(\frac{\delta v_{k+1}-\delta 
v_{k}}{\Delta_{x}}\right)^{2}
\nonumber \\
&=& \int dx\; n(x)\left(\frac{d }{dx}\; \delta v(x)\right)^{2},
\eea
with $n(x)=\lim_{\Delta_{x}\to 0}{\rm 
Re}\left[\frac{\rho_{kk+1}}{\Delta_{x}}\right]\geq 0$ the continuum 
density. For all situations of physical interest the integrand 
may vanish at most in a set of zero measure and is otherwise 
larger than zero. Thus, in the continuum case the above equation is an {\em 
absurdum}, in agreement with the Runge-Gross theorem.

\subsection{TDCDFT on a lattice}
\label{tdcdft}

The problems in extending the Runge-Gross theorem to lattice systems sketched 
in the previous Section can be circumvented by using TD Peierls 
phases instead of TD on-site potentials \cite{StefanucciPerfettoCini:10}.
For arbitrary TD electromagnetic fields described by a scalar and 
vector potential $\tilde{v}(\vr,t)$ and $\tilde{\vA}(\vr,t)$
one can always perform a gauge transformation such that $\tilde{v}(\vr,t)\to
v(\vr)$ and $\tilde{\vA}(\vr,t)\to\vA(\vr,t)$, so that the 
time-dependence resides only in the vector potential.
In particular, for those situations when there is only a TD potential 
$\tilde{v}(\vr,t)$, 
we could always gauge $\tilde{v}$ away and work with a TD vector potential $\vA(\vr,t)$.
In the localized basis we choose, therefore, 
the time-dependent Hamiltonian to have the form
\be
\hat{H}(t) = \hat{K}(t) + \hat{H}_{\rm int},
\label{td_hamil}
\ee
where
\be
\hat{K}(t) = \sum_{m,n}\sum_{\sigma} T_{mn} e^{i \gamma_{mn}(t)} \hat{c}_{m \sigma}^{\dagger} 
\hat{c}_{n \sigma},
\label{kin_op}
\ee
and the time-dependence only resides in the Peierls phases 
$\gamma_{mn}(t)=-\gamma_{nm}(t)$. In a grid basis with grid points 
$\vr_{m}$, the 
on-site matrix elements $T_{mm}=v(\vr_{m})$ contain information on the on-site 
electrostatic potential $v(\vr)$ while the phases 
$\gamma_{mn}(t)=\frac{1}{c}\int_{\vr_{m}}^{\vr_{n}}d{\bf l}\cdot{\bf 
A}(\vr,t)$ describe the 
effects of an external TD vector potential $\vA(\vr,t)$. 

From Eq. (\ref{heseq}) the density at site $m$ satisfies the equation of motion 
\be
\frac{\rm d}{{\rm d} t} n_m(t) = \sum_n J_{mn}(t) + i \left\langle 
[ \hat{H}_{\rm int},\hat{n}_m ] \right\rangle 
\label{eom_dens}
\ee
where we have defined the bond-current operator $\hat{J}_{mn}(t)$ 
according to
\be
\hat{J}_{mn}(t) = \frac{1}{i}\sum_{\sigma} \left( T_{mn} e^{i \gamma_{mn}(t)} 
\hat{c}^{\dagger}_{m\sigma} \hat{c}_{n\sigma} -{\rm  H.c.}  \right) \; .
\ee
In turn, the equation of motion for $J_{mn}(t)$ reads
\be
\frac{\rm d}{{\rm d}t} J_{mn}(t) = K_{mn}(t) \frac{\rm d}{{\rm d}t} 
\gamma_{mn}(t) + F_{mn}(t) 
\ee
with the ``kinetic energy'' density operator
\be
\hat{K}_{mn}(t) = \sum_{\sigma} \left( T_{mn} e^{i \gamma_{mn}(t)} 
\hat{c}_{m \sigma}^{\dagger} \hat{c}_{n \sigma} + {\rm H.c.} \right) 
\ee
and
\be
\hat{F}_{mn}(t) = i  [\hat{H}(t),\hat{J}_{mn}(t)] \; .
\ee

We now ask the question whether the bond-currents $J_{mn}(t)$ along 
those bonds with $T_{mn}\neq 0$ can be 
reproduced by  another Hamiltonian $\hat{H}'(t)$ with a {\em different} 
interaction and a {\em different} TD electromagnetic field, i.e., 
\be
\hat{H}'(t) = \hat{K}'(t) + \hat{H}'_{\rm int}
\label{td_hamil_prime}
\ee
with
\be
\hat{K}'(t) = \sum_{m,n} T_{mn} e^{i \gamma'_{mn}(t)} \hat{c}_{m \sigma}^{\dagger} 
\hat{c}_{n \sigma} \; .
\label{kin_op_prime}
\ee
Note that the matrix elements $T_{mn}$ in Eq.~(\ref{kin_op_prime}) are the 
same as in Eq.~(\ref{kin_op}). Let us define the initial 
configuration of the primed system as the couple 
$\{|\Psi_{0}'\rangle,\g'(0)\}$, that consists of  
the initial state and the
initial value of the Peierls phases.
For the current $J'_{mn}(t)$ of the 
primed system to be the same as the current $J_{mn}(t)$ at the 
initial time $t=0$, 
the initial configuration must be {\em compatible}, i.e., it must fulfill
\be
\langle\Psi'_{0}|\frac{1}{i}\sum_{\sigma} \left( T_{mn} e^{i 
\gamma'_{mn}(0)} 
\hat{c}^{\dagger}_{m\sigma} \hat{c}_{n\sigma} -{\rm  H.c.}  
\right)|\Psi'_{0}\rangle=J_{mn}(0).
\ee
Assuming the existence of at least one compatible configuration we 
demonstrate below that there exist TD Peierls phases $\{\gamma'(t)\}$ for 
which $J_{mn}(t)=J'_{mn}(t)$ for all bonds with $T_{mn}(0)\neq 0$, and 
that these phases are unique.

The bond current densities $J_{mn}(t)$ and $J'_{mn}(t)$ of the two systems 
are the same provided that
\be
K'_{mn}(t) \frac{\rm d}{{\rm d}t} \gamma'_{mn}(t) = 
K_{mn}(t) \frac{\rm d}{{\rm d}t} \gamma_{mn}(t) + F_{mn}(t) - F'_{mn}(t) \; .
\label{equal_currents}
\ee
To proceed further we adopt the same strategy of Vignale \cite{Vignale:04} 
in the context of TD current DFT, and expand all quantities in 
Eq.~(\ref{equal_currents}) in a Taylor series around $t=0$, 
e.g., 
\be
K'_{mn}(t) = \sum_{l=0}^{\infty} K_{mn}'^{(l)} \; t^l \; .
\ee
Inserting the Taylor expansions into Eq.~(\ref{equal_currents}) and equating 
the coefficients with power $l$ then gives
\bea
\lefteqn{
K_{mn}'^{(0)} (l+1) \gamma_{mn}'^{(l+1)} = 
- \sum_{k=0}^{l-1} (k+1) K_{mn}'^{(l-k)} \gamma_{mn}'^{(k+1)} }\nn\\
&& + \sum_{k=0}^{l} (k+1) K_{mn}^{(l-k)} \gamma_{mn}^{(k+1)} + F_{mn}^{(l)} 
- F_{mn}'^{(l)} \; .
\label{recursive}
\eea
The quantities $K'_{mn}(t)$ and $F'_{mn}(t)$ depend explicitly 
on the phases $\{\gamma'(t)\}$ and therefore their $l$-th derivative 
depend on all the $k$-th derivatives $\{\gamma'^{(k)}\}$ with $k\leq l$.
Thus, under the mild condition that the initial compatible 
configuration is chosen in such a way that
\be 
K'_{mn}(0)=K_{mn}'^{(0)}\neq 0 \;, 
\ee
Eq. (\ref{recursive}) constitute a system of recursive relations to 
determine all Taylor coefficients of $\gamma'_{mn}(t)$ for all bonds 
with $T_{mn}\neq 0$. 

A direct corollary of this result is that if the interaction Hamiltonian 
$\hat{H}_{\rm int}'=\hat{H}_{\rm int}$ 
and the initial configuration 
$\{|\Psi_{0}'\rangle,\gamma'(0)\}=\{|\Psi_{0}\rangle,\gamma(0)\}$, then the Peierls 
phases $\gamma'(t)=\gamma(t)$, i.e., for any fixed initial 
configuration there is a one-to-one correspondence between the 
bond-currents and the Peierls phases. Consequently, we can think of
the TD many-body state $|\Psi(t)\rangle$, and hence of all TD 
expectation values, as a functional of the bond-currents: 
$|\Psi(t)\rangle=|\Psi[J](t)\rangle$. Another important corollary 
is that the bond-currents of a system with 
interaction $\hat{H}_{\rm int}$ can be reproduced in a system of 
noninteracting electrons, $\hat{H}'_{\rm int} =0$, which we call the 
KS system. These two corollaries lay down the foundations 
of a TD bond-current functional theory (BCFT) for discrete (or lattice) 
Hamiltonians, where the basic variable is the bond currents $J(t)$
and the basic KS field is the Peierls phases $\g(t)$.

Once the equality of the bond currents  is established, 
it follows from Eq.~(\ref{eom_dens}) that if 
\be
[ \hat{H}_{\rm int},\hat{n}_m] = [ \hat{H}'_{\rm int},\hat{n}_m] = 0 
\label{int_dens_commute}
\ee
the time evolution of the densities $n_m(t)$ and $n'_m(t)$ 
is also the same, provided that they are the same at the initial time, 
$n_m(0)=n'_m(0)$. 
It is worth noting that for  
Hubbard-like interactions, Eq.~(\ref{int_dens_commute}) is satisfied. 

Finally we would like to emphasize that in the above proof the possibility of 
reproducing the bond-currents in a system with a different interaction heavily 
relies on the convergence of the Taylor series, which has here simply been  
assumed \cite{note2}. 
An alternative proof based on the Picard-Lindel\"of theorem 
for non-linear differential 
equations has been recently proposed by Tokatly \cite{Tokatly}; 
this proof has the advantage of avoiding the Taylor expansions but it 
remains unpredictive on the maximum time until which a solution exist.
Another merit of Tokatly's 
stategy is that TDBCFT can be generalized to primed systems 
with different hopping integrals $T'_{mn}\neq T_{mn}$.

\section{Model Hamiltonian for time-dependent transport}
\label{modhamsec}

Tight-binding based model systems of interacting electrons are 
often used to study transport through nanostructures such as quantum dots. 
Although most of these studies are restricted to the steady-state regime, more 
recently there has been increasing activity to describe the time evolution 
towards the steady state as the system is driven out of equilibrium 
by applying a bias in the leads. These studies use a range of methods such as, 
e.g., TDDFT \cite{KurthStefanucciAlmbladhRubioGross:05,BurkeCarGebauer:05,BokesCorsettiGodby:08,ZhouChu:09,ZhengChenMoKooTianYamYan:10}, 
generalized master equations 
\cite{MoldoveanuManolescuTangGudmundsson:10}, many-body perturbation theory 
\cite{MyohanenStanStefanucciLeeuwen:08,MyohanenStanStefanucciLeeuwen:09,vonFriesenVerdozziAlmbladh:10}, 
the time-dependent density-matrix renormalization group 
\cite{HeidrichMeisnerFeiguinDagotto:09,HeidrichMeisnerGonzalezHassaniehFeiguinDagotto:10,BranschaedelSchneiderSchmitteckert:10}, 
a quantum trajectory approach \cite{Oriols:07}, 
or real-time path integral \cite{MuehlbacherRabani:08} 
and Monte Carlo approaches \cite{WernerOkaEcksteinMillis:10}. 

Here we will apply the TDBCFT formalism developed in the previous Section to 
describe 
TD quantum transport through a one-dimensional Anderson-like model 
system \cite{Anderson:61}. We will consider a two-terminal setup where two 
semi-infinite non-interacting leads are coupled to an interacting single-level 
impurity. The total Hamiltonian of the system reads
\be
\hat{H}(t) = \hat{H}_{\rm C}(t) + \hat{H}_{\rm L}(t) + \hat{H}_{\rm R}(t) + 
\hat{H}_{\rm T} .
\label{hamil_int}
\ee
We model the Hamiltonian for the impurity as
\be
\hat{H}_{\rm C}(t) = \sum_{\sigma} \varepsilon_{C}(t) 
\hat{d}_{\sigma}^{\dagger} \hat{d}_{\sigma} + \hat{H}_{\rm int} ,
\label{hamil_dot}
\ee
with a Hubbard-like electron-electron interaction
\be
\hat{H}_{\rm int} = U \hat{d}_{\uparrow}^{\dagger} \hat{d}_{\uparrow} 
\hat{d}_{\downarrow}^{\dagger} \hat{d}_{\downarrow} .
\label{interaction}
\ee
The Hamiltonian for the semi-infinite lead $\alpha=L,R$ is taken of 
the form
\bea
\hat{H}_{\alpha}(t) &=& \sum_{i=1}^{\infty}\sum_{\sigma} \left( \varepsilon_\alpha + 
W_{\alpha}(t) \right) 
\hat{c}_{i\sigma\alpha}^{\dagger} \hat{c}_{i\sigma\alpha} \nn \\
&& - \sum_{i=1}^{\infty}\sum_{\sigma} \left( V_{\alpha} \hat{c}_{i\sigma\alpha}^{\dagger} 
\hat{c}_{i+1\sigma\alpha} + {\rm H.c.} \right) ,
\label{hamil_leads}
\eea
and the tunneling Hamiltonian 
\be
\hat{H}_{T} = - \sum_{\sigma}\sum_{\alpha=L,R} \left( 
V_{\rm link}\hat{d}_{\sigma}^{\dagger} \hat{c}_{1\sigma \alpha}   
+ {\rm H.c.} \right) 
\label{hamil_tunn}
\ee
connects the impurity to both left and right leads. 

We note that the Hubbard interaction (\ref{interaction}) satisfies 
Eq.~(\ref{int_dens_commute}) since it 
commutes with the on-site density 
$\hat{n}_{\sigma}= \hat{d}_{\sigma}^{\dagger} \hat{d}_{\sigma}$ in the 
impurity as well as with the on-site density 
$\hat{n}_{m\sigma\alpha} = \hat{c}_{m \sigma \alpha}^{\dagger} 
\hat{c}_{m \sigma \alpha}$ in lead $\alpha$, i.e, 
\be
[ \hat{H}_{\rm int}, \hat{n}_{\sigma}] = [ \hat{H}_{\rm int}, 
\hat{n}_{m\sigma\alpha}] = 0 \; .
\ee
In our model the time-dependence appears exclusively 
through a TD on-site potential, that is the bias $W_{\alpha}(t)$ in 
lead $\alpha$ and the on-site energy $\varepsilon_C(t)$ at  
the impurity. Therefore, the gauge transformation 
$\hat{c}_{m \sigma\alpha} \to \hat{c}_{m \sigma\alpha}
e^{i \beta_{\alpha}(t)}$ with 
$\beta_{\alpha}(t)= \int_0^t {\rm d} t'\, W_{\alpha}(t')$ and 
$\hat{d}_{\sigma} \to \hat{d}_{\sigma} e^{i \beta_C(t)}$ 
with $\beta_C(t)= \int_0^t {\rm d} t' \, \varepsilon_C(t')$
leads to Peierls phases of the particular form $\gamma_{m\alpha n\alpha}(t) = 0$ 
for any pair of sites $m$ and $n$ in lead $\alpha$ and to 
\be
\gamma_{\rm L}(t) = \int_0^t {\rm d} t' 
\left( W_{\rm L}(t') - \varepsilon_C(t') \right)
\label{phase_left}
\ee
and 
\be
\gamma_{\rm R}(t) = \int_0^t {\rm d} t' 
\left( \varepsilon_C(t') - W_{\rm L}(t') \right)
\label{phase_right}
\ee
for the Peierls phases of the bonds connecting the left and right leads to the 
impurity. Under this gauge transformation the TD on-site potential 
is gauged away and the Hamiltonian (\ref{hamil_int}) 
is transformed into an Hamiltonian of the form in Eqs. 
(\ref{td_hamil}-\ref{kin_op})
\be
\hat{H}(t) = \tilde{\hat{H}}_{\rm C} + \tilde{\hat{H}}_{\rm L}
+ \tilde{\hat{H}}_{\rm R} + \tilde{\hat{H}}_{\rm T}(t) .
\label{hamil_int_gauge}
\ee
The transformed impurity Hamiltonian $\tilde{\hat{H}}_{\rm 
C}=\hat{H}_{\rm int}$ and lead Hamiltonians 
\be
\tilde{\hat{H}}_{\alpha} = \sum_{i=1}^{\infty}\sum_{\sigma} 
\varepsilon_{\alpha} 
\hat{c}_{i\sigma\alpha}^{\dagger} \hat{c}_{i\sigma\alpha} 
- \sum_{i=1}^{\infty} \sum_{\sigma} \left( V_{\alpha} \hat{c}_{i\sigma\alpha}^{\dagger} 
\hat{c}_{i+1\sigma\alpha} + {\rm H.c.} \right)
\ee
become independent of time whereas  
the transformed tunneling Hamiltonian 
\be
\tilde{\hat{H}}_{T}(t) = - \sum_{\sigma} \sum_{\alpha=L,R} \left( 
 V_{\rm link} e^{i \gamma_{\alpha}(t)} \hat{d}_{\sigma}^{\dagger} 
\hat{c}_{1\sigma \alpha} +{\rm  H.c.} \right)  
\ee
carries all the time dependence.

\subsection{Local Peierls phase approximation and the ABALDA functional}

\begin{figure}[t]
\includegraphics[width=0.47\textwidth]{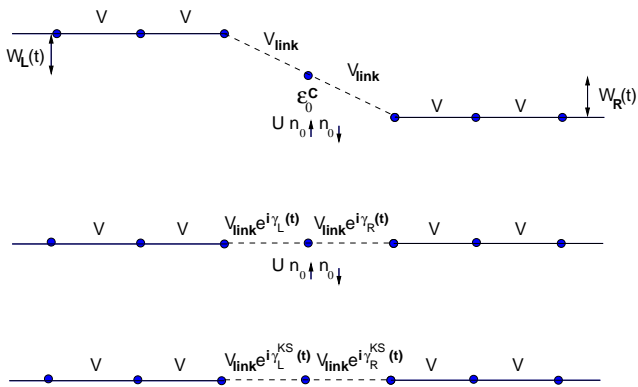}
\caption{Schematic representation of the model system. The biased system (top) 
with Hubbard-type interaction on the dot is gauge-equivalent to the system 
(middle) with time-dependent Peierls phases $\gamma_L(t)$ and $\gamma_R(t)$ 
(see Eqs.~(\protect\ref{phase_left}) and (\protect\ref{phase_right})). In the 
local Peierls phase approximation, the Kohn-Sham system (bottom) is assumed 
to have nonvanishing Peierls phases (Eqs.~(\protect\ref{ks_phase_left}) and 
(\protect\ref{ks_phase_right})) only at the same links as the interaction 
system (middle). 
}
\label{sys_sketch}
\end{figure}

The TDBCFT presented in Section \ref{tdcdft} guarantees that the 
bond-currents $J_{mn}$ of the interacting system can be reproduced in a 
non-interacting KS system with TD Peierls phases 
$\gamma^{\rm KS}_{mn}(t)$. 
In the following we make a local approximation for the $\gamma^{\rm 
KS}_{mn}(t)$: we set them to  zero except 
for those bonds where the external Peierls phases $\gamma_{mn}(t)$ are 
non-zero,  i.e., the 
bonds connecting the impurity to the leads, see Fig.~\ref{sys_sketch}. 
Within this approximation, we write the only two KS Peierls phases similarly to Eqs. 
(\ref{phase_left}-\ref{phase_right})
\be
\gamma_{\rm L}^{\rm KS}(t) = \int_0^t {\rm d} t' 
\left( W_{\rm L}(t') - v_{\rm KS}(t') \right),
\label{ks_phase_left}
\ee
and 
\be
\gamma_{\rm R}^{\rm KS}(t) = \int_0^t {\rm d} t' 
\left( v_{\rm KS}(t') - W_{\rm L}(t') \right) ,
\label{ks_phase_right}
\ee
where $\varepsilon_{C}(t)$ is replaced by the on-site KS potential 
$v_{\rm KS}(t)$ that we write as the sum of the external on-site 
potential $\varepsilon_{C}(t)$, the Harteee potential $\frac{1}{2} U 
n(t)$ (with $n(t)$ the TD density at the impurity)
and the exchange-correlation (XC) potential $v_{\rm xc}(t)$
\be
v_{\rm KS}(t) = \varepsilon_0^{\rm C}(t) + \frac{1}{2} U n_0(t) + v_{\rm xc}(t) 
\; .
\label{kspot}
\ee
Note that by a gauge transformation we could go back to the original 
gauge in which there is an on-site potential instead of the Peierls phases. 
In the original gauge the KS system is described by the Hamiltonian
\be
\hat{H}(t) = \hat{H}_{\rm C}^{\rm KS}(t) + \hat{H}_{\rm L}(t) + 
\hat{H}_{\rm R}(t) + \hat{H}_{\rm T} ,
\ee
where the lead and the tunneling Hamiltonians, $\hat{H}_{\alpha}(t)$ and 
$\hat{H}_T$, are unchanged compared to the interacting Hamiltonian 
(i.e., they are given by Eqs.~(\ref{hamil_leads}) and (\ref{hamil_tunn}), 
respectively) and the KS Hamiltonian for the impurity is
\be
\hat{H}_{\rm C}^{\rm KS}(t) = \sum_{\sigma} v_{\rm KS}(t) 
\hat{d}_{\sigma}^{\dagger} \hat{d}_{\sigma} .
\label{ks_hamilton}
\ee

We still need to specify the approximation used for the XC
potential $v_{\rm xc}(t)$. In this work we use an 
adiabatic version of the local density approximation (LDA) for the static, 
non-uniform Hubbard model. The construction of this functional 
\cite{LimaSilvaOliveiraCapelle:03} follows a similar strategy as the one used 
to construct the usual LDA based on the model of the uniform electron gas. The 
crucial difference is that the underlying model is the uniform Hubbard model 
whose exact solution can be constructed via the Bethe ansatz. Just as the 
XC energy of the uniform electron gas serves as input for the usual LDA, 
the exact XC energy per site of the uniform Hubbard model then serves as input 
for the Bethe-ansatz LDA (BALDA) used for non-uniform Hubbard models
\cite{LimaSilvaOliveiraCapelle:03,SilvaLimaMalvezziCapelle:05}. In the context 
of TDDFT, an adiabatic version of this functional (adiabatic 
BALDA, ABALDA) local in both space and time, e.g., $v_{\rm xc}[n](i,t) = 
v_{\rm xc}(n_i(t))$, has been suggested by Verdozzi \cite{Verdozzi:08}. 
The form of the BALDA XC potential derives from a parametrization of 
the XC 
energy per particle of the uniform Hubbard model. This parametrization deviates 
somewhat from the exact, numerical XC energy of the Hubbard model 
\cite{XianlongPoliniTosiCampoCapelleRigol:06,AkandeSanvito:10}, especially for 
weak interactions, but here we are not concerned about these differences. The 
crucial property for our purposes is the existence of a derivative 
discontinuity at half filling \cite{LimaOliveiraCapelle:02} (see discussion 
below). 

\begin{figure}[t]
\includegraphics[width=0.47\textwidth]{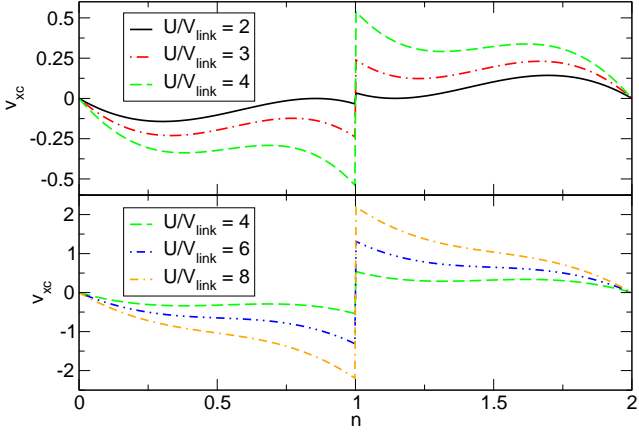}
\caption{Smoothened BALDA exchange-correlation potential 
(Eq.~(\protect\ref{vxc_smooth})) as function of the density for various 
values of $U/V_{\rm link}$ with smoothing parameter $a=10^{-4}$.}
\label{vxc_balda_fig}
\end{figure}

The original BALDA has been proposed for a system which not only has the same 
on-site energy but also the same interaction for all sites and 
also the hopping connecting neighbouring sites is the same throughout the 
chain. In Ref.~\onlinecite{SilvaLimaMalvezziCapelle:05} the BALDA functional 
has already been used successfully for {\em site-dependent interactions}. It 
should be noted that this is a deviation from the usual LDA philosophy 
where the XC energy of the uniform gas is used for a non-uniform system but 
the interaction is unchanged. Here we also take the same approach and use 
the BALDA for a system with interactions only at one site. In addition, we 
are interested in situations when the interacting region is weakly connected 
to two non-interacting leads, i.e., the hopping $V$ within the leads may be 
different from the coupling $V_{\rm link}$ of the leads to the interacting 
region. For this situation a generalization of the original BALDA 
functional has been suggested \cite{KurthStefanucciKhosraviVerdozziGross:10}.
The explicit form of this modified BALDA XC potential reads 
\be
v_{\rm xc}[n]=\theta(1-n)v^{(<)}_{\rm xc}[n]-\theta(n-1)v^{(<)}_{\rm 
xc}[2-n],
\ee
where
\be
v^{(<)}_{\rm xc}[n]=
-\frac{1}{2} {U} n - 2 V _{\rm link}\left[
\cos\left(\frac{\pi n}{2}\right) -
\cos\left(\frac{\pi n}{\zeta}\right) \right] \; .
\label{vxc_balda_1}
\ee
The parameter $\zeta$ is determined by the equation 
\be
\frac{2 \zeta}{\pi} \sin(\pi/\zeta) = 4 \int_0^{\infty} {\rm d} x \,
\frac{J_0(x) J_1(x)}{x [1 + \exp({U} x/(2 V_{\rm link}))]}
\label{eq_zeta}
\ee
where $J_{i=0,1}(x)$ are Bessel functions. The BALDA XC potential has a 
discontinuous jump at half-filling \cite{LimaOliveiraCapelle:02}: 
\be
v_{\rm xc}(n=1^+) - v_{\rm xc}(n=1^-) = U - 4 V_{\rm link} 
\cos\left(\frac{\pi}{\zeta}
\right) \; .
\label{discont}
\ee
Here it is interesting to note that in the limit of very weak coupling 
($V_{\rm link} \to 0$) the discontinuity just reduces to $U$ which is the 
charging energy required to put a second electron on the interacting 
impurity if it is already occupied by one electron, i.e., in this limit 
the functional becomes exact. Although the exact XC potential for the 
Hubbard model is certainly discontinuous, in our situation where a single 
interacting impurity is coupled to two non-interacting leads, the coupling 
to the leads introduces some broadening in the levels of the isolated 
impurity and it is reasonable to expect that this broadening leads to 
a smoothening of the discontinuity. We therefore introduce a smoothening 
in our model and use the following XC potential
\be
\tilde{v}_{\rm xc}[n]=f(n) v^{(<)}_{\rm xc}[n]-(1-f(n)) v^{(<)}_{\rm 
xc}[2-n],
\label{vxc_smooth}
\ee
where 
\be
f(n) = \frac{1}{\exp((n-1)/a)+1}
\ee
and $a$ is a smoothening parameter. In Fig.~\ref{vxc_balda_fig} we show the 
BALDA XC potential for different values of $U/V_{\rm link}$ using the value 
$a=10^{-4}$ for the smoothening parameter. 

It has recently been shown \cite{KurthStefanucciKhosraviVerdozziGross:10}  that 
this discontinuity is crucial to describe Coulomb blockade. Moreover, in a 
time-dependent picture and in the parameter regime of Coulomb blockade it also 
prevents the biased system to reach a steady state. More details on these 
findings will be discussed below. 

\subsection{Time propagation with embedding} 

By mapping the interacting problem onto a non-interacting one we have already 
achieved a considerable simplification. However, we are still dealing with an  
infinitely extended system which has to be treated numerically. This can be 
achieved by an embedding technique which maps the infinite system exactly 
onto a tractable, finite problem. 

In the localized site basis the KS Hamiltonian (\ref{ks_hamilton}) has the 
matrix structure
\be
\vH = \left(
\begin{array}{ccc}
\vH_{LL} & \vH_{LC} & 0 \\
\vH_{CL} & \vH_{CC} & \vH_{CR} \\
0 & \vH_{RC} & \vH_{RR} \\
\end{array} \right).
\label{hamil_matrix}
\ee
It is possible to derive the equation of motion for the $k$-th KS
single-particle orbital projected onto the central region, 
$\psi_{k,C}(t)$,
which reads
\bea
\left[ i\frac{d}{dt} - \vH_{CC}(t) \right] \psi_{k,C}(t) =
\int_0^t {\rm d} t' \; \vSi_{\rm em}(t,t') \psi_{k,C}(t')
\nn\\
+ \sum_{\alpha} \vH_{C \alpha} \vg_{\alpha \alpha}(t,0) 
\psi_{k,\alpha}(0),
\label{psic_eom}
\eea
where
\be
\vSi_{\rm em}(t,t') = \sum_{\alpha=L,R} \vH_{C \alpha} \vg_{\alpha \alpha}(t,t') 
\vH_{\alpha C}
\label{sigma_emb}
\ee
is the retarded embedding self energy. Here, the retarded Green's function 
$\vg_{\alpha \alpha}(t,t')$ of the isolated lead $\alpha$ satisfies the equation 
of motion 
\be
\left[ i \frac{d}{dt} - \vH_{\alpha\alpha}(t) \right] \vg_{\alpha \alpha}(t,t') =
\delta(t-t') \; 
\label{galpha}
\ee
with boundary conditions $\vg_{\alpha \alpha}(t,t^{+})=0$ and 
$\vg_{\alpha \alpha}(t,t^{-})=-i$.
The TD density at the impurity can be calculated from the 
KS orbitals as in Eq. (\ref{tddens}), i.e.,
\be
n(t) = \sum_k^{\rm occ} | \psi_{k,C}(t)|^2 \; ,
\ee
where the sum runs over the occupied KS states. 
In the present work we use the algorithm described in 
Ref.~\onlinecite{KurthStefanucciAlmbladhRubioGross:05} to propagate 
Eq.~(\ref{psic_eom}) starting from the self-consistent BALDA ground state of 
the coupled lead-dot-lead system. 

\section{The derivative discontinuity and its connection to Coulomb blockade}
\label{CBsec}

\subsection{Ground state and non-equilibrium steady state}

We will use the KS Hamiltonian in the BALDA 
approximation to describe transport for our model of a single interacting 
impurity connected to two non-interacting tight-binding leads. This model has 
been studied extensively in the literature, and the results from 
other methods  can be used for a validation 
of our KS treatment. In the present Section we first assess the quality of our approximate 
BALDA functional in the ground state by comparing to recent Quantum 
Monte Carlo (QMC)
results for the same model system. Then we look into the nonequilibrium steady 
state for the system exposed to a DC bias in the leads, assuming that such a 
steady state exists. 

Using the techniques of non-equilibrium Green's functions one can derive a 
self-consistency condition for the ground or steady-state electron density 
$n^{\infty}$ at the impurity  which reads
\be
n^{\infty } = 2 \sum_{\alpha={\rm L,R}} \int_{-\infty}^{\varepsilon_F+W_{\alpha}}
\frac{{\rm d}\omega}{2 \pi} \Gamma(\omega-W_{\alpha}) |G(\omega)|^2  \; .
\label{dens_sc}
\ee
Here, $W_{\alpha}$ ($\alpha={\rm L,R}$) is the constant bias applied in lead $\alpha$
(of course, the ground state density is obtained for $W_{\rm 
L}=W_{\rm R}=0$) 
and
\be
G(\omega) = \left(\omega - v_{\rm KS}[n^{\infty}] - \sum_{\alpha={\rm L,R}} 
\Sigma_{\alpha}(\omega - W_{\alpha}) \right)^{-1} 
\ee
is the retarded Green's function in frequency space at the impurity. 
$\Sigma_{\alpha}(\omega)$ 
is the Fourier transform of the $\alpha$-lead contribution to the embedding 
self energy (\ref{sigma_emb}) which, for the present case 
of non-interacting tight-binding leads with vanishing onsite energies, 
$\varepsilon_{\alpha}=0$, is given explicitly by
\bea
\lefteqn{
\Sigma_{\alpha}(\omega) = \Lambda_{\alpha}(\omega) - \frac{i}{2}
\Gamma_{\alpha}(\omega) } \nn \\
&=& \frac{V_{\rm link}^2}{2 V_{\alpha}^2}
\left\{
\begin{array}{ll}
\omega_{\alpha} - \sqrt{\omega_{\alpha}^2-4 V_{\alpha}^2} &
\mbox{\hspace*{4mm}for $\omega_{\alpha} > 2 V_{\alpha}$} \\
\omega_{\alpha} + \sqrt{\omega_{\alpha}^2-4 V_{\alpha}^2} &
\mbox{\hspace*{4mm}for $\omega_{\alpha} < - 2 V_{\alpha}$} \\
\omega_{\alpha} - i\sqrt{4 V_{\alpha}^2 - \omega_{\alpha}^2} &
\mbox{\hspace*{4mm}for $|\omega_{\alpha}| \leq 2 V_{\alpha}$}
\end{array}\right.\;\;\;
\label{self_emb}
\eea
with real and imaginary parts $\Lambda_{\alpha}(\omega)$ and 
$\Gamma_{\alpha}(\omega)$, respectively. Finally, the total width 
function is $\Gamma(\omega) = \sum_{\alpha} \Gamma_{\alpha}(\omega)$ and 
$\varepsilon_F$ is the Fermi energy of the contacted system in the ground 
state. 

\begin{figure}[t]
\includegraphics[width=0.47\textwidth]{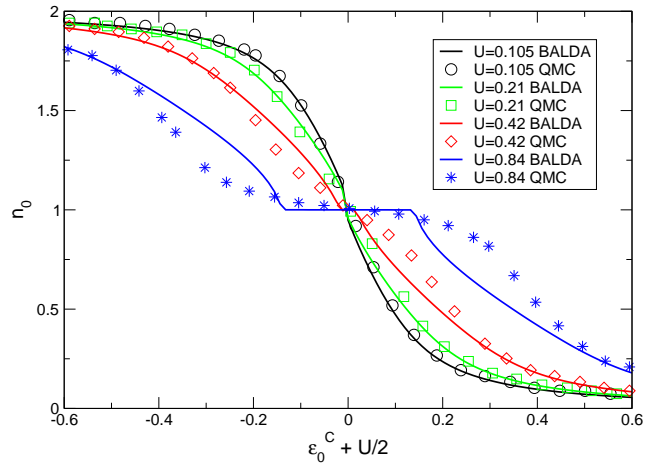}
\caption{Comparison of BALDA and Quantum Monte Carlo (QMC) ground state 
densities at the impurity as function of onsite energy $\varepsilon_0^C$ 
for different values of the interaction. The coupling between leads and 
impurity is $V_{\rm link}=0.1803$. All energies are given in units of the 
hopping $V$ in the leads. The QMC results were extracted from 
Ref.~\protect\onlinecite{WangSpataruHybertsenMillis:08}. }
\label{gsdens_gate}
\end{figure}

We have first solved Eq.~(\ref{dens_sc}) for the ground state density 
$n_0:= n^{\infty}$ for $W_{\rm L}=W_{\rm R}=0$ and compared to recent QMC
calculations presented by Wang et al. in 
Ref.~\onlinecite{WangSpataruHybertsenMillis:08}. Here we focus on the 
dependence of $n_0$ on the on-site energy $\varepsilon_C$ at the 
impurity. 
We set $\varepsilon_F=0$ (half filling of the leads) and for the Hubbard 
interaction we use $U=0.105,0.21,0.42,0.84$ (all energies are given in units 
of the equal hopping in left and right leads $V_{\rm L}=V_{\rm R}=V$). 
For the hopping between leads and impurity we take a weak $V_{\rm link}=0.1803$. 
In order to achieve a meaningful comparison to the QMC data, the value of our 
$V_{\rm link}$ is a factor of $1/\sqrt{2}$ smaller than the one used in 
Ref.~\onlinecite{WangSpataruHybertsenMillis:08} since Wang et al. 
considered an impurity coupled to a single lead only \cite{verdozzi_comm}. 
For the smoothening parameter we choose the value $a=10^{-4}$. 
Fig.~\ref{gsdens_gate} shows the ground state density as function of on-site 
energy $\varepsilon_C$. We see that the QMC and BALDA results agree 
surprisingly well, especially for weak interactions. For  
strong interactions the BALDA develops a clear Coulomb blockade step, i.e., 
the density hardly changes over a significant range of onsite energies. 
Although the step in BALDA extends over a smaller range of onsite energies 
than in QMC, the agreement is still quite reasonable. 
Already at this stage it is clear that the feature which gives rise to 
the Coulomb blockade step is the derivative discontinuity built into the 
BALDA functional. 

Now we turn our attention to the biased, non-equilibrium situation. 
If we {\em assume} that the system evolves toward a steady state in the 
long time limit then the steady-state value of the density at the 
impurity is given by the self-consistent solution of Eq. (\ref{dens_sc}). 
We solve the latter
for different values of the bias $W_{\rm L}$ applied in the left lead. 
The system parameters are $\varepsilon_C=2$, $U=2$, $\varepsilon_F=1.5$ 
and $a=10^{-4}$. 

\begin{figure}[t]
\includegraphics[width=0.47\textwidth]{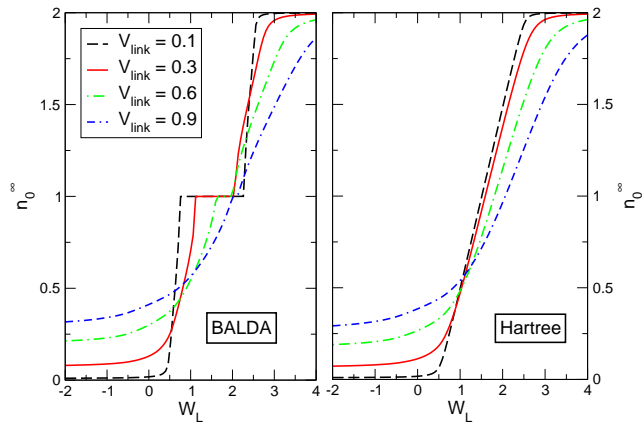}
\caption{Steady state density as function of the applied bias $W_L$ for 
different values of $V_{\rm link}$. Left panel: BALDA steady state densities,
right panel: Hartree steady state densities. }
\label{steady}
\end{figure}

In the left panel of Fig.~\ref{steady} we show the steady state density as function of the applied bias for different values of the hopping $V_{\rm link}$ 
between leads and impurity. We clearly see a plateau in the steady-state 
density at unity, i.e., when the impurity is occupied by exactly 
one electron. This 
plateau becomes wider and more step-like as $V_{\rm link}$ decreases which 
corresponds to the usual picture of Coulomb blockade: the dot can  only be 
occupied by zero, one, or two electrons, and the energy cost for double 
occupancy is given by the Hubbard interaction $U$. It is important to 
emphasize here that in the plateau region it is crucial that we used an 
XC potential with a smoothened discontinuity. If one uses the truly 
discontinuous XC potential, the steady-state condition actually does not 
have any solution at all in this parameter region, indicating that the initial 
steady-state assumption was not justified. This point will be further 
investigated in the next Section. 

In the right panel of Fig.~\ref{steady} we show the steady-state densities for 
the same parameter range if one uses the Hartree approximation, i.e., when 
$v_{\rm xc}$ in Eq.~(\ref{kspot}) is set to zero. One clearly sees that in this 
case the step in the steady-state density is completely absent. 
Instead, for small $V_{\rm link}$ the steady-state density rises almost 
linearly within a certain bias range. For larger values of $V_{\rm link}$, the 
Hartree and BALDA steady-state densities are qualitatively quite similar. 

\subsection{Time-dependent transport}

In the previous Section we {\em assumed} that under application of a DC bias 
the system evolves towards a steady state for which we then solved a 
self-consistency condition to obtain the steady-state density. In the present 
Section we will show that for certain parameters the steady-state assumption 
is actually {\em not always} justified: when perturbing the system, which is 
initially in its ground state, by switching on a DC bias  the 
time evolution does {\em not necessarily} lead to a steady state but 
rather to an oscillatory state of dynamical density oscillations.

\begin{figure}[t]
\includegraphics[width=0.47\textwidth]{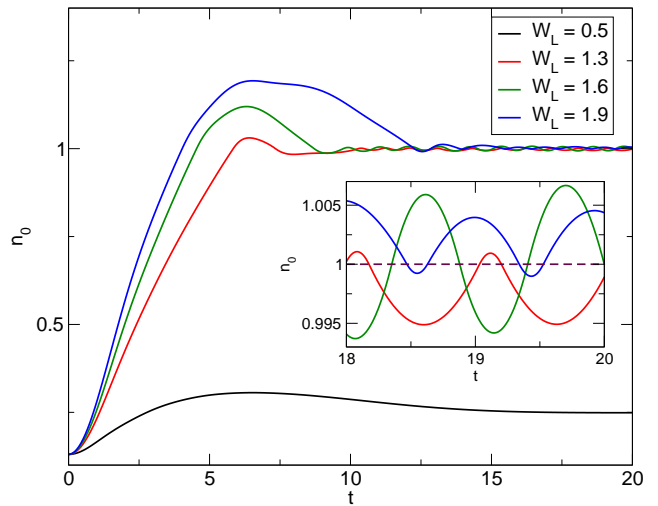}
\caption{TD density in BALDA for different values of the bias 
$W_L$ for $V_{\rm link}=0.3$. Inset: Magnification of the TD density at 
large times. The straight line at unity is a guide for the eye.}
\label{tddens_fig}
\end{figure}

In Fig.~\ref{tddens_fig} we show the time evolution of the 
quantum-dot density  from its intial, ground-state value when, at $t=0$, a DC bias is 
suddenly switched on in the left lead, i.e., $W_{\rm L}(t) = 
\theta(t) W_{\rm L}$ with $W_{\rm L}=0.5,1.3,1.6,1.9$. The 
parameters are the same as those used for Fig.~\ref{steady} with $V_{\rm link}=0.3$. 
According to the steady-state picture of 
Fig.~\ref{steady}, the first value $W_L=0.5$ corresponds to the rising flank 
of the density as function of bias while the other biases correspond to the 
plateau. The TD density corresponding to bias 
$W_{\rm L}=0.5$ can be seen to evolve smoothly towards its steady state value. 
However, looking closely at the evolution of the density for the other 
bias values (inset of Fig.~\ref{tddens_fig}) one can see that 
for these cases the system does not evolve towards a steady state but rather 
approaches a dynamical state of periodically oscillating density. 
Furthermore, the density oscillates around unity, i.e., single occupancy 
of the impurity. The amplitude 
of these oscillations is rather small, of the order of $5 \times 10^{-3}$. 
We also note that the larger the value of the bias (in the range of the 
plateau of Fig.~\ref{steady}), the larger the fraction of time for which 
the density exceeds the critical value of unity. 

\begin{figure}[t]
\includegraphics[width=0.47\textwidth]{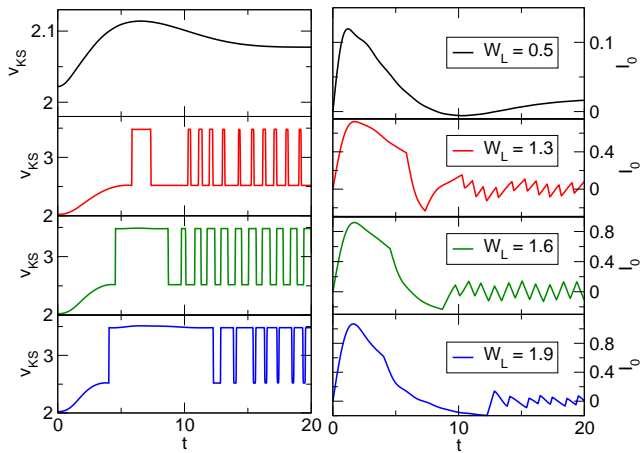}
\caption{Time-dependent KS potentials and currents at the impurity for 
different values of the bias $W_L$ for $V_{\rm link}=0.3$. Left panels: 
BALDA KS potentials, right panels: TD currents at the impurity site. }
\label{td_kspot_curr}
\end{figure}

The qualitative difference of the response of the system for biases 
within and outside the step region of Fig.~\ref{steady} becomes quite 
apparent when looking at the time-dependent KS potentials (left panels of 
Fig.~\ref{td_kspot_curr}) and currents through the impurity (right panels 
of the same figure). While for the bias outside the step region 
($W_{\rm L}=0.5$) 
both quantities behave  smoothly, for the other biases the KS potentials 
show rapid variations which are due to the discontinuity at $n=1$: as long 
as the time-dependent density stays below or above unity, the KS potential 
changes smoothly in time. However, when the density crosses unity, the KS 
potential jumps discontinuously (or very rapidly for the smoothened KS 
potentials used here). The rapid variation of the KS potential allows 
us to understand why for these cases a steady state is not achieved: 
consider the situation when the time-dependent density crosses the critical 
value of unity from below. At the instant $t_{c1}$ when the jump in the KS 
potential occurs, the rate of change of the density is positive, 
$\dot{n}(t_{c1})>0$, and immediately after $t_{c1}$ the density continues to grow 
because the maximum rate of change of the density is limited by the inertia 
of the electrons. However, for times right after $t_{c1}$ the 
significant increase of the KS potential tends to push down the density, i.e., 
$\ddot{n}(t)<0$, and therefore $n(t)$ will cross the 
critical value of unity from above at some time $t_{c2}>t_{c1}$. 
At that time $\dot{n}(t_{c2})<0$, the KS potential will 
suddenly be lowered and electrons will be attracted by the quantum 
dot, i.e., $\ddot{n}(t)>0$. Therefore, the density will eventually 
crosses unity at some time $t_{c3}>t_{c2}$ from below and the cycle 
will start over.

The above results have been obtained with an XC potential with a smoothened 
discontinuity for which, in principle, the condition for the steady-state 
density discussed in the previous Section does have a solution. This solution 
is not reached by the time propagation because the rate of change of the  
density is limited by $V_{\rm link}$ \cite{Verdozzi:08}. The rate of change of the KS 
potential when the density crosses the (smoothened) discontinuity is 
instead limited by the smoothening parameter $a^{-1}$. Therefore the steady state should not be attained 
whenever the rate of change of the KS potential is larger than the largest 
possible rate of change of the density. 

The time-dependent currents in the Coulomb blockade regime (lower three 
right panels of Fig.~\ref{td_kspot_curr}) show sawtooth-like oscillations at 
the impurity site. 
By the equation of motion for the currents, this is consistent with the fact 
that the KS potentials form a train of almost rectangular potential steps. 
In contrast, the current away from the Coulomb blockade 
regime (upper right panel of Fig.~\ref{td_kspot_curr}) evolves smoothly 
towards its steady state value. This is qualitatively similar to the behaviour 
which is found for time-dependent Hartree calculations (in any parameter 
regime). Finally, we would like to point out that the oscillations just 
described are entirely due to electron correlations and are therefore 
different from oscillations occuring due to the presence of 
single-particle bound states 
\cite{Stefanucci:07,KhosraviKurthStefanucciGross:08,KhosraviStefanucciKurthGross:09}. 

\section{Conclusions}
\label{concsec}

In the present work we have described the difficulties in generalizing the 
fundamental Runge-Gross proof of time-dependent density functional theory to 
systems described by a lattice Hamiltonian. To overcome these difficulties we 
have employed the time-dependent bond current instead of the density as basic 
variable. We were then able to prove a
one-to-one correspondence between time-dependent bond currents 
and Peierls phases describing the electromagnetic field on the lattice 
and therefore proposed a time-dependent bond current functional theory as the 
proper extension of TDDFT to lattice systems.

In a second part we have described a time-dependent approach to electron 
transport in model systems using this TDBCFT. We proposed a local Peierls 
phase approximation assuming that the Peierls phases of the noninteracting 
KS system are nonvanishing only for the same links as for the interacting case. 
A simple model of an interacting impurity connected to two noninteracting leads 
has been studied employing a functional which exhibits an explicit 
derivative discontinuity leading to a 
time-dependent KS potential which jumps discontinuously 
as the particle number on the impurity crosses an integer 
\cite{LeinKuemmel:05}. It has 
been demonstrated that this discontinuity is (i) crucial to describe Coulomb 
blockade and (ii) prevents the system from reaching a steady state 
by time-evolution out of the initial ground state upon application of a bias 
in the leads. Instead we find that after some transient time the system 
reaches a dynamical state of undamped density oscillations. 
In this picture, Coulomb blockade manifests itself as a 
dynamic phenomenon of sequentially charging and discharging the impurity.

Here we demonstrated the crucial role played by the discontinuity in the 
XC potential to properly describe Coulomb blockade, a physical phenomenon 
which is due to electronic correlations, for a simple model system. 
This model is not easily generalized to 
more complicated situations such as, e.g., more orbitals or degrees of freedom 
per site or more complicated lattices. As for the latter case, there have 
recently been activities to construct the XC energy per particle which also 
display a derivative discontinuity for more complicated model systems such 
as a graphene-like hexagonal Hubbard lattice \cite{IjaesHarju:10} or the 
Hubbard model in three dimensions \cite{KarlssonPriviteraVerdozzi}. However, 
the challenge remains to construct practical and universal approximations 
with a derivative discontinuity not only for lattice systems but also for 
systems described by continuum Hamiltonians. 

\acknowledgments
We would like to acknowledge discussions with Elham Khosravi, 
Claudio Verdozzi, Hardy Gross, Ilya Tokatly, Enrico Perfetto, and Michele 
Cini. S.~K.~ acknowledges funding by the "Grupos Consolidados UPV/EHU del 
Gobierno Vasco" (IT-319-07) and the European Community's Seventh Framework 
Programme (FP7/2007-2013) under grant agreement No. 211956.


\begin{thebibliography}{10}
\bibitem{note2}  
For the case of TD current DFT for continuum Hamiltonians, 
the convergence issue for physical systems has been discussed in 
Ref.~\onlinecite{Vignale:04}.
\end{thebibliography}
\end{document}